\documentclass[doublespacing,reviewcopy]{elsarticle}
\usepackage{graphicx}
\usepackage{makeidx}
\usepackage{amsmath,amsfonts,dsfont}
\usepackage[english]{babel}
\usepackage{tikz,pgflibraryshapes,xcolor,xkeyval}
\usepackage[]{hyperref}
\usepackage{subfigure}
\usepackage{setspace}

\newcommand{\FN}{FitzHugh-Nagumo\ }
\newcommand{\HR}{Hindmarsh-Rose\ }
\newcommand{\HH}{Hodgkin-Huxley\ }
\newcommand{\ML}{Morris-Lecar\ }
\newcommand{\IF}{Integrate-and-Fire\ }

\newcommand{\Real}{\mathds{R}}

\newcommand{\la}{\leftarrow}

\newcommand{\ra}{\rightarrow}
\newcommand{\Ra}{\Rightarrow}
\newcommand{\Acal}{\mathcal{A}}
\newcommand{\Bcal}{\mathcal{B}}
\newcommand{\Ccal}{\mathcal{C}}
\newcommand{\Dcal}{\mathcal{D}}

\newcommand{\Wcal}{\mathcal{W}}
\newcommand{\Xcal}{\mathcal{X}}

\newcommand{\col}[1]{\mathrm{col}\left(#1\right)}
\newcommand{\sgn}[1]{\mathrm{\, sign}\left(#1\right)}

\newcommand{\ben}{\begin{enumerate}}
\newcommand{\een}{\end{enumerate}}

\newcommand{\beq}{\begin{equation}}
\newcommand{\eeq}{\end{equation}}

\newcommand{\beqa}{\begin{equation}\begin{array}{l}}
\newcommand{\eeqa}{\end{array}\end{equation}}

\newcommand{\beqm}{\begin{equation}\begin{array}{ll}}
\newcommand{\eeqm}{\end{array}\end{equation}}

\newcommand{\splt}[1]{\begin{split}#1\end{split}}

\newcommand{\bbm}{\begin{bmatrix}}
\newcommand{\ebm}{\end{bmatrix}}

\newcommand{\bpm}{\begin{pmatrix}}
\newcommand{\epm}{\end{pmatrix}}

\newcommand{\bit}{\begin{itemize}}
\newcommand{\eit}{\end{itemize}}

\newcommand{\barr}{\begin{array}}
\newcommand{\earr}{\end{array}}

\newcommand{\abs}[1]{\left|#1\right|}
\newcommand{\norm}[1]{\left\Vert#1\right\Vert}
\newcommand{\brac}[1]{\left(#1\right)}
\newcommand{\cbrac}[1]{\left\{#1\right\}}

\newenvironment{proof}[1][Proof]{\begin{trivlist}
\item[\hskip \labelsep {\bfseries #1}]}{\end{trivlist}}

\newtheorem{thm}{Theorem}
\newtheorem{lem}{Lemma}
\newtheorem{prop}{Proposition}[section]
\newtheorem{defn}{Definition}[section]
\newtheorem{exam}{Example}[section]
\newtheorem{remark}{Remark}[section]

\graphicspath{{figures/}}

\begin{document}

\begin{frontmatter}

\title{Semi-passivity and synchronization of diffusively coupled neuronal oscillators}

\author[TUe]{Erik Steur}
\ead{e.steur@tue.nl}
\author[LE,RI]{ Ivan Tyukin}
\ead{I.Tyukin@le.ac.uk}
\author[TUe]{Henk Nijmeijer}
\ead{h.nijmeijer@tue.nl}

\address[TUe]{Dept. of Mechanical Engineering, Eindhoven University of Technology, P.O. Box 513 5600 MB,  Eindhoven, The Netherlands}
\address[LE]{Department of Mathematics, University of Leicester, University Road, Leicester, LE1 7RH, UK}
\address[RI]{Laboratory for Perceptual Dynamics, RIKEN BSI, Wako-shi, Saitama, Japan}

\begin{abstract}
We discuss synchronization in networks of neuronal oscillators which are interconnected via diffusive coupling, i.e. linearly coupled via gap junctions. In particular, we present sufficient conditions for synchronization in these networks using the theory of semi-passive and passive systems. We show that the conductance-based neuronal models of Hodgkin-Huxley, Morris-Lecar, and the popular reduced models of FitzHugh-Nagumo and Hindmarsh-Rose all satisfy a semi-passivity property, i.e. that is the state trajectories of such a model remain oscillatory but bounded provided that the supplied (electrical) energy is bounded. As a result, for a wide range of coupling configurations, networks of these oscillators are guaranteed to possess ultimately bounded solutions. Moreover, we demonstrate that when the coupling is strong enough the oscillators become synchronized. Our theoretical conclusions are confirmed by computer simulations with coupled \HR and \ML oscillators. Finally we discuss possible ``instabilities'' in networks of oscillators induced by the diffusive coupling.
\end{abstract}

\end{frontmatter}

\section{Introduction}
Synchronous behavior is witnessed in a variety of biological systems. Examples include the simultaneous flashing of fireflies and crickets that are chirping in unison \cite{str1993}, the synchronous activity of pacemaker cells in the heart \cite{pes1975} and synchronized bursts of individual pancreatic $\beta$-cells \cite{she1988}. For more examples see \cite{pik2003,str2003} and the references therein. It is well known that individual neurons in parts of the brain discharge their action potentials in synchrony. In fact, synchronous oscillations of neurons have been reported in the olfactory bulb, the visual cortex, the hippocampus and in the motor cortex \cite{gra1994,sin1999}. Presence or absence of synchrony in the brain is often linked to specific brain function or critical physiological state (e.g. epilepsy). Hence, understanding conditions that will lead to such behavior, exploring the possibilities to manipulate these conditions, and describe them rigorously is vital for further progress in neuroscience and related branches of physics.

We present results on synchronization of ensembles of neuronal oscillators which are being interconnected via gap-junctions, i.e. a linear electrical coupling of the form $g\cdot(V_1(t)-V_2(t))$ where the constant $g$ represents the synaptic conductance and $V_1(t)-V_2(t)$ denotes the difference in membrane potential of the neurons at the pre-synaptic side and the post-synaptic side at time $t$, respectively. Recently it has been pointed out that gap-junctions play an important role in synchronization of individual neurons \cite{ben2004}.



Several attempts have been made to understand when synchronization of neurons coupled via gap junctions occurs. In \cite{erm2009,lew2003,man2007,nom2003,vel2003} phase equations and phase response curves are used to analyse neurons coupled via gap junctions. They all conclude that for increasing coupling the synchronous state becomes stable. However, the use of phase equations is only justified when the coupling between the cells is weak. In general, the results for strong coupling are rare \cite{coo2008}. In \cite{coo2008} Coombes uses a piecewise linear model of spiking neurons which allows to extend the results for weak coupling (using phase equations) for strong coupling. Chow and Kopell \cite{cho2000} used \IF kind of models to investigate synchronization via gap junctions. They showed using spike response functions (for the \IF models an analytic expression for this function exists) that, depending on the shape of the spikes, the firing frequency and the coupling strength, stable phase locked states exist. When the coupling is large the oscillators will synchronize. They showed using simulations that for more realistic models the results hold true as well, however no rigorous mathematical proof is presented. In \cite{lab2003} conditions for synchrony in two coupled \HH neurons are presented; If the coupling between the neurons is strong enough, then the neurons will synchronize. In \cite{oud2004} synchronization for multiple interconnected chaotic \HR neurons is discussed.

We will generalize the results obtained in \cite{oud2004} and present conditions for synchrony of diffusively coupled identical neuronal oscillators for general network topology.
From the zoo of models of neuronal activity (see \cite{izh2004} for a review) we will focuss on four popular oscillators, namely the conductance based, biophysically meaningful models of \HH \cite{hod1952} and \ML\cite{mor1981}, and the more abstract models derived by \FN \cite{fit1961,nag1962} and \HR\cite{hin1984}.
First we demonstrate that, despite the difference in the range of behavior that these models are capable to produce, these models have an important collective property. This property is that each model is semi-passive\footnote{we will formally introduce semi-passivity in Definition \ref{def:semi-passivity} in Section \ref{sec:semi-passivity-general}.}.
Second, using the concept of semi-passivity, introduced in \cite{pog1998}, we will show that a set of these diffusively coupled neuronal oscillators will always possess bounded solutions. Next, under condition that the coupling between the neurons is large enough, i.e. there is a high-conductive pathway between the neurons, we show that the oscillators will become synchronized.

This paper is organized as follows. In Section \ref{sec:semi-passivity-general} we introduce the notion of semi-passivity and we show that four models mentioned above are all semi-passive.
Next, in Section \ref{sec:semi-passivity-appl}, a theorem adopted from \cite{pog2001} is presented which provides sufficient conditions under which the oscillators show synchronous behavior. We demonstrate in Section \ref{sec:examples} using computer simulations that ensembles of \HR and \ML oscillators will end up in synchrony whenever the coupling between the neurons is large enough.
In Section \ref{sec:instabilities} we briefly discuss that it is not obvious that systems being interconnected via diffusive coupling will have bounded solutions and eventually end up in synchrony. In particular, we show that two ``dead'' cells can become ``alive'' when being interconnected via diffusive coupling, i.e. the cells start to oscillate due to the interaction. Finally, Section \ref{sec:discussion} concludes of the paper.

Throughout this paper we use the following notations.
The symbol $\Real$ stands, as usual, for the real numbers, $\Real_+$ denotes the following subset of $\Real$: $\Real_+ = \cbrac{x \in \Real | x \geq 0}$.
The Euclidian norm in $\Real^n$ is denoted by $\norm{\cdot}$, $\norm{x}^2 = x^\top x$ where the symbol $^\top$ stands for transposition. The symbol $I_n$ defines the $n \times n$ identity matrix and the notation $\col{x_1, \ldots, x_n}$ stands for the column vector containing the elements $x_1, \ldots, x_n$. A function $V : \Real^n \ra \Real_+$ is called positive definite if $V(x) > 0$ for all $x \in \Real^n \setminus \cbrac{0}$. It is radially unbounded if $V(x) \ra \infty$ if $\norm{x} \ra \infty$. If the quadratic form $x^\top P x$ with a symmetric matrix $P = P^\top$ is positive definite, then the matrix $P$ is positive definite, denoted as $P>0$. The
symbol $\Ccal^r$ denotes the space of functions that are at least $r$ times differentiable. Consider $k$ interconnected systems and let $x_j$ denote the state of a single system, then the systems are called synchronized if $\lim_{t\ra\infty}\norm{x_i(t)-x_j(t)}=0$, $i,j \in\{1,2,\ldots,k\}$.

\section{Semi-passivity}\label{sec:semi-passivity-general}
We represent a neuronal oscillator as the general system
\beq\label{eq:system_pas}
    \splt{
        \dot{x} &=  f(x)+B u,\\
        y &= C x,
    }
\eeq
where state $x \in \Real^n$, input $u\in\Real$ is an depolarizing or hyperpolarizing (input) current and output $y\in\Real$ denotes the membrane potential of the neuron. Furthermore, $f:\Real^n\ra\Real^n$ is a $\Ccal^1$-smooth vector field and the matrices $B$ and $C$ are of appropriate dimensions.

\begin{defn}[Passivity and semi-passivity \cite{wil1972,pog2001}]\label{def:semi-passivity}
The system \eqref{eq:system_pas} is called
\begin{itemize}
\item[i)] passive in $\Dcal\subset \Real^n$ if there exists a nonnegative function $V:\Dcal \ra \Real_+$, $\Dcal$ is open, connected and invariant under the dynamics \eqref{eq:system_pas}, $V(0) = 0$, such that the following dissipation inequality
\begin{equation}\label{eq:passive}
\dot{V}(x) = \frac{\partial V(x)}{\partial x}\left(f(x)+B u\right)\leq y^\top u
\end{equation}
holds; if $\Dcal = \Real^n$ the system is called passive;\\

 \item[ii)] semi-passive in $\Dcal$ if there exists a nonnegative function $V:\Dcal \subset \Real^n \ra \Real_+$, $\Dcal$ is open, connected and invariant under \eqref{eq:system_pas}, $V(0) = 0$, such that
%
\beq\label{eq:semi-passivity}
    \dot{V}(x) = \frac{\partial V(x)}{\partial x}\left(f(x)+B u\right)\leq y^\top u -H(x),
\eeq
where the function $H:\Dcal\subset \Real^n\ra\Real$ is nonnegative outside the ball $\Bcal$ with radius $\rho$
\[
    \exists \rho > 0 , \ \norm{x}\geq\rho \Ra H(x)\geq \varrho\brac{\norm{x}},
\]
with some nonnegative continuous function $\varrho(\cdot)$ defined for all $\norm{x}\geq\rho$; if $\Dcal = \Real^n$ the system is called semi-passive;\\

 \item[iii)] strictly semi-passive (in $\Dcal$) if the function $H(\cdot)$ is positive outside some ball $\Bcal\subset\Dcal$.
\end{itemize}
\end{defn}

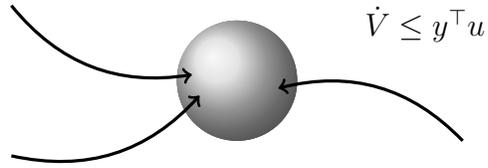
\begin{figure}[B!]
    \begin{center}
            \begin{tikzpicture}
            \centering
                \draw (-5,-1.5) node {};
                \draw (-5,1.5) node {};
                \draw (5,-1.5) node {};
                \draw (5,1.5) node {};
                \shade[ball color=gray!30] (0,0) circle (0.8cm);
                \draw (2.5,0.75) node {\large $\dot{V}\leq y^\top u$};
                \draw [overlay,->,very thick](-3,1) to[bend right] (-0.60,0.08);
                \draw [overlay,->,very thick](-3,-1) to[bend right] (-0.50,-0.2);
                \draw [overlay,->,very thick](3,-0.8) to[bend right] (0.55,-0.1);
            \end{tikzpicture}
            \caption{Semi-passivity; every solution enters the ball $\norm{x}\leq\rho$ in finite time and stays there as time increases.}
        \label{fig:passivity}
    \end{center}
\end{figure}

A semi-passive system behaves similar to a passive system for large enough $\norm{x}$. Hence a semi-passive system that is interconnected by a feedback $u = \varphi(y)$ satisfying $y^\top \varphi(y) \leq 0$ has ultimately bounded solutions \cite{wil1972,pog2001}, i.e. regardless how the initial conditions are chosen, every solution of the closed-loop system enters a compact set in a finite time and stays there, see Figure \ref{fig:passivity}. Moreover, this compact set does not depend on the choice of initial conditions.

Consider $k$ identical neuronal oscillators of the form
\beq\label{eq:systems_sync}
    \splt{
        \dot{x}_j &=  f(x_j)+B u_j,\\
        y_j &= C x_j,
    }
\eeq
where $j = 1,\ldots,k$ denotes the number of each system in the network, $x_j \in \Real^n$ the state, $u_j \in \Real$ the input and $y_j \in \Real$ the output of the $j^\mathrm{th}$ system, i.e. the membrane potential, smooth vector field $f:\Real^n \ra \Real^n$ and vectors $B=[ 1 \ 0 \ \ldots \ 0 ]^\top$ and $C=[ 1 \ 0 \ \ldots \ 0]$ are of appropriate dimensions. Note that many neuronal models are in this form or can be put in this form via a well-defined change of coordinates.

The $k$ neurons \eqref{eq:systems_sync} are coupled via \emph{diffusive} coupling, i.e. a mutual interconnection through linear output coupling of the form
\beq \label{eq:diffuse_coupling}
    u_j = - \gamma_{j1}\brac{y_j - y_1} - \gamma_{j2}\brac{y_j - y_2} - \ldots - \gamma_{jk}\brac{y_j - y_k}
\eeq
where $\gamma_{ji} = \gamma_{ij} \geq 0$ represents the synaptic conductance and $y_i-y_j$ is the difference in membrane potential of neurons $i$ and $j$.

Defining the $k \times k$ \emph{coupling matrix} as
\beq
    \Gamma =
    \bbm
        \sum_{j = 2}^{k} \gamma_{1j} & -\gamma_{12} & \ldots & -\gamma_{1k}\\
        -\gamma_{21} & \sum_{j = 1, j\neq 2}^{k} \gamma_{2j}  & \ldots & -\gamma_{2k}\\
        \vdots & \vdots & \ddots & \vdots\\
        -\gamma_{k1} & -\gamma_{k2} & \ldots  & \sum_{j = 1}^{k-1} \gamma_{kj}
    \ebm
\eeq
the diffusive coupling functions \eqref{eq:diffuse_coupling} can be written as
\beq \label{eq:diffuse_coupling2}
    u = - \Gamma y
\eeq
where $u = \col{u_1,\ldots, u_k}$ and $y = \col{y_1,\ldots, y_k}$.
Since $\Gamma = \Gamma^\top$ all its eigenvalues are real and $\Gamma$ is singular because all rowsums equal zero. Moreover, applying Gerschgorin's theorem (cf. \cite{stew1990}) about the localization of the eigenvalues, it is easy to verify that $\Gamma$ is positive semi-definite. We assume that the network cannot be divided into two or more disconnected networks. Hence the matrix $\Gamma$ has a simple zero eigenvalue.

\begin{prop}\label{prop:bounded}
Consider a network of $k$ diffusively coupled systems \eqref{eq:systems_sync}, \eqref{eq:diffuse_coupling}. Assume that each system in the network is semi-passive, then the solutions of all connected systems in the network are ultimately bounded.
\end{prop}

\begin{proof}
The proof is adopted from \cite{pog2001}.
Let the $j^\mathrm{th}$ system in the network be semi-passive with the storage function $V(x_j)$, where $x_j$ is the state of the $j^\mathrm{th}$ system. Denote $W(x)=\sum_{j=1}^k V(x_j)$ where $x = \col{x_1,\ldots,x_k}$, then
\beq\label{eq:dotW}
    \dot{W}(x) = \sum\limits_{j=1}^k \dot{V}(x_j) \leq \sum\limits_{j=1}^k y_j^\top u_j - H(x_j) = -y^\top \Gamma y - \sum\limits_{j=1}^k H(x_j)\leq 0,
\eeq
outside some ball in $\Real^{nk}$.
Note that the quadratic term $y^\top \Gamma y$ is nonnegative since $\Gamma$ is semi-positive definite.
This directly implies that the solutions of the interconnected systems are bounded and exist for all $t \geq t_0$.
\end{proof}

\begin{remark}
Even if the systems are not identical, but each individual system is semi-passive, then the network will still have bounded solutions. This follows directly from \eqref{eq:dotW}, i.e. the storage function for the network is simply the sum of the storage functions of the individual oscillators.
\end{remark}

\begin{remark}
Consider a collection of $k$ neurons that interact via chemical synapses, where the chemical synapse is modeled as the Fast Threshold Modulation (FTM) coupling introduced in \cite{som1993}, i.e.
\beq
    u_{i} = \sum \limits_{j=1}^k -g_{ij} H(y_j-\theta) (\alpha - y_i),
\eeq
where $g_{ij}\in \Real_{>0}$ denotes the synaptic conductance, $\alpha\in\Real$ is the synaptic reversal potential which determines whether the synapse is inhibitory or excitatory, and the function $H(\cdot)$ is typically chosen as the Heaviside function such that neuron $j$ will influence neuron $i$ only if the membrane potential of neuron $j$ exceeds some threshold $\theta\in\Real$. It is not hard to verify that semi-passive neuronal oscillators interconnected via chemical synapses have bounded solutions. (This follows from the fact that $\sum y_i u_i \leq 0$ outside some ball in $\Real^k$, i.e. the ``supplied energy'' is bounded.)
\end{remark}

We are now ready to prove that the neuronal models of Hodgkin-Huxley, Morris-Lecar, FitzHugh-Nagumo and Hindmarsh-Rose all satisfy the semi-passive property. Hence the solutions of networks of these oscillators with a diffusive coupling exist and are bounded.

\subsubsection*{\HH model}
The most important model in computational neuroscience is probably the \HH model \cite{hod1952}.
Consider the \HH equations :
\beq\label{eq:HH}
    \splt{
        C\dot{x}_1 &= {g}_{Na} x_2^3 x_3 \brac{E_{Na} - x_1} + {g}_{K} x_4^4 \brac{E_{K} - x_1} + g_L \brac{E_{L} - x_1} + I + u \\
        \dot{x}_i &= \alpha_{i}(x_1)\brac{1-{x_i}} - \beta_{i}(x_1) {x_i},\qquad i = 2,3,4 \\
    }
\eeq
with $y=x_1$ is the membrane potential, state $x \in \Xcal \subset \Real^4$, input $u\in \Real$, positive constants ${g}_{Na}, {g}_K, g_L, C \in \Real$ and constants $I, E_{Na}, E_K, E_{L}\in \Real$. The functions $\alpha_j(\cdot)$ and $\beta_j(\cdot)$ are defined as
\beq
    \splt{
        \alpha_{2}(s)&=\frac{25-s}{10\brac{e^{\brac{2.5-s/10 }}-1}}, \\
        \alpha_{3}(s)&=0.07e^{-s/20}, \\
        \alpha_{4}(s)&=\frac{10-s}{100\brac{e^{\brac{1-{s/10}}}-1}},\\
        \beta_{2}(s)&=4e^{-s/18}, \\
        \beta_{3}(s)&=\frac{1}{e^{\brac{3-{s/10}}}+1},\\
        \beta_{4}(s)&=0.125e^{-s/80}. \\
    }
\eeq
The states $x_i$ represent so-called activation particles which satisfy $x_i(t)\in (0,\ 1)$ for all $t\geq t_0$ whenever $x_i(t_0)\in (0,\ 1)$.

\begin{prop}
The \HH model is semi-passive in $\Xcal$ where
\beq
    \Xcal = \{x\in\Real^4 \vert 0< {x_i}< 1,\ i = 2,3,4\}.
\eeq
\end{prop}

\begin{proof}

First, we will prove that for all $t_0\leq t_1$, $t_0,t_1\in\Real$:
\ben
    \item[C1)]\label{claim:A} $x_1(t)$ exists on the interval $t \in [t_0,\ t_1]$ and remains bounded if the input $u$ is bounded;
    \item[C2)]\label{claim:B} $x_i(t)\in (0,1)$ on the interval $t \in [t_0,\ t_1]$ provided $x_i(t_0) \in (0,\ 1)$.
\een

We do so by invoking a contradiction argument. Suppose that C1) does not hold. Let us denote
\beq
u^\ast = \sup_{t\in[t_0,t_1]}\|u(t)\|.
\eeq
According to assumptions of the proposition such $u^\ast$ must exist. The right-hand side of  (\ref{eq:HH}) is locally Lipschitz, hence its solutions are defined over a finite time interval. Let $[t_0,T]$ be the maximal interval of their existence. Let us pick some arbitrarily large constant $M\in\Real_+$. Then there should exist a time instant $t_1'$ such that
\beq\label{eq:insert:0}
\|x(t)\|\geq M, \quad \forall \ t\geq t_1'.
\eeq
Consider the internal dynamics
\beq
    \dot{x}_i = \alpha_{i}(x_1)\brac{1-{x_i}} - \beta_{i}(x_1) {x_i},\qquad i = 2,3,4.
\eeq
One can easily verify that $\alpha_i(x_1) > 0$, $\beta_i(x_1)>0$ for all (bounded) $x_1$. Hence on the boundary $x_i=0$ we have $\dot{x}_i>0$ and at the boundary $x_i=1$ we have $\dot{x}_i < 0$, i.e. $x_i$ can not cross the boundaries. Hence the set $(0,1)$ is forward invariant under the $x_i$ dynamics, i.e. for all $x_i(t_0) \in (0,\ 1)$,
\beq\label{eq:insert:1}
0 < x_i(t) < 1, \quad \forall \ t\in[t_0,T].
\eeq
Then, according to (\ref{eq:insert:1}), (\ref{eq:HH}) the following holds
\beq\label{eq:insert:2}
\|x(t)\|\leq e^{-\lambda (t-t_0)}|x_1(t_0)|+ \rho + \frac{1}{\lambda} u^\ast, \  \ \forall \ t\in[t_0,T]
\eeq
where $\rho$, $\lambda$ are positive constants of which the value do not depend on $M$. Combining (\ref{eq:insert:0}) and (\ref{eq:insert:2}) we obtain
\beq
M\leq \|x(t)\|\leq e^{-\lambda (t-t_0)}|x_1(t_0)|+ \rho + \frac{1}{\lambda} u^\ast, \  \ \forall \ t\in[t_1',T]
\eeq
where $M$ is arbitrarily large and $\rho$, $x_1(t_0)$, and $1/\lambda u^\ast$ are fixed and bounded. Hence we have reached contradiction, and C1) hold. This automatically implies that C2) holds too.

To finalize the proof of semi-passivity of (\ref{eq:HH}), consider the storage function $V:\Xcal \ra \Real_+$, $V = \frac{1}{2 C}x_1^2 + \tfrac{1}{2} \sum \limits_{i=2}^4 x_i^2$. Then
\beq\label{eq:HH_dV}
    \splt{
    \dot{V} =& x_1 u \\
    &-\brac{ {g}_{Na} x_2^3 x_3 + g_K x_4^4 + g_L} x_1^2\\
    &+\brac{ {g}_{Na} x_2^3 x_3 E_{Na}+ g_K x_4^4E_K + g_L E_L + I} x_1\\
    &- \sum \limits_{i=2}^4 \brac{ \alpha_{i}(x_1)\brac{\brac{x_i-\tfrac{1}{2}}^2 - \tfrac{1}{4}} + \beta_{i}(x_1) {x_i}^2}.
    }
\eeq
Note that $- \brac{ \alpha_{i}(x_1)\brac{\brac{x_i-\tfrac{1}{2}}^2 - \tfrac{1}{4}} + \beta_{i}(x_1) {x_i}^2} \leq 0$ for each $x_i$ outside $(0,\ 1)$. Because C2) holds we obtain
\beq\label{eq:HH_dV_ineq}
    \splt{
    \dot{V} \leq& x_1 u -g_L x_1^2+ c_1x_1 \\
                &- \sum \limits_{i=2}^4 \brac{ \alpha_{i}(x_1)\brac{\brac{x_i-\tfrac{1}{2}}^2 - \tfrac{1}{4}} + \beta_{i}(x_1) {x_i}^2 }
   }
\eeq
where constant
\beq
\splt{
c_1 =\max \limits_{d_1,d_2\in [0,1]}&\abs{ d_1{g}_{Na}  E_{Na}+ d_2 g_K E_K + g_L E_L + I} \times \\
&\times\sgn{ d_1{g}_{Na}  E_{Na}+ d_2 g_K E_K + g_L E_L + I}.
}
\eeq
Given that \eqref{eq:HH_dV_ineq} holds for all $t$, the \HH model is semi-passive in $\Xcal$.
\end{proof}

\subsubsection*{\ML model}
The \ML model \cite{mor1981} is a planar system that models the voltage oscillations in the barnacle giant muscle fiber.
The \ML model is given by the following equations
\beq
    \splt{
        C\dot{x}_1 &= g_L\brac{E_L-x_1} + g_{Ca}\alpha_\infty\brac{x_1}\brac{E_{Ca}-x_1}+g_K x_2 \brac{E_K-x_1}+ I + u,\\
        \dot{x}_2 &= \eta\brac{x_1}\brac{\beta_\infty(x_1) - x_2},
    }
\eeq
with $y=x_1$ denoting the membrane potential, state $x \in \Xcal \subset \Real^2$, input $u\in \Real$, constant parameters $E_L,\ E_{Ca},\ E_K\in\Real$, positive constants $g_L,\ g_{Ca},\ g_K\in \Real$ and functions
\beq
    \splt{
        \alpha_\infty(s) &= \frac{1}{2}\brac{1+\tanh\brac{\frac{s - E_1}{E_2}}}, \\
        \beta_\infty(s) &= \frac{1}{2}\brac{1+\tanh\brac{\frac{s - E_3}{E_4}}}, \\
        \eta(s) &= \bar{\eta}\cosh\brac{\frac{s-E_3}{2 E_4}},
    }
\eeq
with $\bar{\eta}>0$, $E_1,E_2,E_3,E_4,\bar{\eta} \in \Real$. Like in the \HH equations, the states $x_2$ represent an activation particle which satisfies $x_2(t)\in (0,\ 1)$ for all $t\geq t_0$ provided $x_2(t_0)\in (0,\ 1)$.

\begin{prop}
The \ML model is semi-passive in $\Xcal$ where
\beq
    \Xcal = \{x\in\Real^2 \vert 0< {x_2} < 1\}.
\eeq
\end{prop}

\begin{proof}
Notice the forward invariance of the set $(0,\ 1)$ under the $x_2$-dynamics. The proof is similar to the proof for the \HH equations.
\end{proof}

\subsubsection*{\FN model}
The \FN model \cite{fit1961,nag1962} is one of the simplest models of the spiking dynamics of a neuron. The model is given by the following set of differential equations
\beq
    \splt{
        \dot{x}_1 &= x_1-\frac{x_1^3}{3} - x_2 + I + u,\\
        \dot{x}_2 &=\phi\brac{ x_1 + a - b x_2 },
    }
\eeq
where $y = x_1$ represents the membrane potential, state $x=(x_1, \ x_2)^\top \in \Real^2$, input $u\in \Real$ and positive constants $a,b,\phi \in \Real$. Constant parameter $I\in\Real$ determines the output-mode of the model (either spiking or quiet).

\begin{prop}
The \FN equations satisfy the semi-passivity property \eqref{eq:semi-passivity}.
\end{prop}

\begin{proof}
Consider the storage function $V:\Real^2 \ra \Real_+$
\beq
V = \frac{1}{2}\brac{x_1^2 + \frac{1}{\phi}x_2^2}.
\eeq
Then
\beq
    \dot{V}  = x_1 u -\frac{x_1^4}{3} + x_1^2 + I x_1 - b x_2^2+a x_2.
\eeq
Therefore $\dot{V}(x_1,x_2) \leq x_1 u - H(x_1,x_2)$ with $H(x_1,x_2) = \frac{x_1^4}{3} - x_1^2 - I x_1 + b x_2^2 - a x_2$, i.e. the \FN neuron is semi-passive.
\end{proof}

\subsubsection*{\HR model}
Consider the \HR \cite{hin1984} equations
\beq
    \splt{
    \dot{x}_1 &= -a x_1^3 + b x_1^2 + x_2 -x_3 + I + u \\
    \dot{x}_2 &= c-d x_1^2 - x_2\\
    \dot{x}_3 &= r\brac{s \brac{x_1 + w}-x_3}
    }
\eeq
where $y = x_1$ represents the membrane potential, state $x=(x_1, \ x_2,\ x_3)^\top \in \Real^3$, input $u\in \Real$ and constant positive parameters $a,b,c,d,r,s,w\in\Real$. The constant parameter $I\in\Real$ determines again the output-mode of the model, which in this case, depending on the choice of parameters, can be resting, bursting or spiking. Moreover, for some parameters it can even behave chaotically.

\begin{prop}
The \HR model is semi-passive.
\end{prop}

\begin{proof}
The proof is adopted from \cite{oud2004}.
Consider the storage function $V:\Real^3 \ra \Real_+$
\beq
V = \tfrac{1}{2}\brac{x_1^2 + \mu x_2^2 + \tfrac{1}{r s} x_3^2}
\eeq
with constant $\mu>0$. Hence
\beq\label{eq:dV_HR}
\dot{V} =  x_1 u -a x_1^4 + b x_1^3 + x_1 x_2 +I x_1 +  \mu c x_2 -\mu d x_1^2 x_2 -\mu x_2^2 + w x_3 - \tfrac{1}{s}x_3^2.
\eeq
Let
\beq
    -a x_1^4 - \mu d x_1^2 x_2 = -a \lambda_1 x_1^4 -a(1-\lambda_1)\brac{x_1^2 + \tfrac{\mu d}{2a(1-\lambda_1)}x_2}^2 + \tfrac{\mu^2 d^2}{4a(1-\lambda_1)}x_2^2
\eeq
and
\beq
    -\mu x_2^2+x_1x_2 = -\mu \lambda_2 x_2^2 - \mu(1-\lambda_2)\brac{x_2 -\tfrac{1}{2 \mu(1-\lambda_2)}x_1}^2 + \tfrac{1}{4 \mu(1-\lambda_2)}x_1^2
\eeq
with $\lambda_i \in (0,1)\subset\Real$, $i=1,2$. Then
\beq
\splt{
\dot{V} =& x_1 u\\
&-a \lambda_1 x_1^4  + b x_1^3 + \tfrac{1}{4 \mu(1-\lambda_2)}x_1^2 +I x_1 \\
&-\brac{\mu \lambda_2 - \tfrac{\mu^2 d^2}{4a(1-\lambda_1)}}x_2^2 + \mu c x_2   \\
&- \tfrac{1}{s}x_3^2 + w x_3\\
&- \mu(1-\lambda_2)\brac{x_2 -\tfrac{1}{2 \mu(1-\lambda_2)}x_1}^2 \\
&-a(1-\lambda_1)\brac{x_1^2 + \tfrac{\mu d}{2a(1-\lambda_1)}x_2}^2.
}
\eeq
Let $\mu < \tfrac{4 a \lambda_2(1-\lambda_1)}{d^2}$. Then it follows directly that the \HR model satisfies the semi-passivity property \eqref{eq:semi-passivity}.
\end{proof}

\begin{remark}
Many biophysically meaningful neuronal models, i.e. conductance based models like the \HH and \ML models, share the same structure, see for instance \cite{hod1952,mor1981,tra1991}.
In particular, the evolution of the membrane potential is given by an equation of the form
\beq
    C \dot{v}(t) = u(t) + \sum \limits_{j=1}^k I_j(t)
\eeq
where $v\in\Real$ denotes the membrane potential, $C\in\Real_{>0}$ is the membrane capacity, $u\in\Real$ is the input and ionic currents $I_j(t) = g_j(t) (E_j - v(t))$ with constant reversal potential $E_j\in\Real$ and time-varying conductance $g_j(t)>0$ for all $t$. The conductance is typically given as
\beq
    g_j = \bar{g}_j \prod \limits_{i=1}^m s_i^{p_{ij}}
\eeq
with maximal conductance $\bar{g}_j \in \Real_{>0}$, nonnegative integers $p_{ij}$ and voltage dependent gating variables $s_i(v(t))$, where the gating variables satisfy $s_i(t) \in (0,\ 1)$ for all $t\geq t_0$ whenever $s_i(t_0) \in (0,\ 1)$.

All models of neuronal oscillators of this form are semi-passive (in $\Real\times(0,\ 1)\times \ldots\times(0,\ 1)$), and the proof for semi-passivity is similar to the proof presented for the \HH model.
\end{remark}

\begin{remark}
Consider the class of \IF neurons, i.e. neuronal models of the form
\beq
    \dot{x} = a - b x + u, \quad \mathrm{if}\ x \geq x_{thres}, \ \mathrm{then}\ x \la c,
\eeq
where the output $y = x$ represents the membrane potential, $u$ is the input, positive constants $a,b$, $x_{thres}$ is the threshold potential and $c$ is the value to which the membrane potential $x$ is reset to after firing. The state of an \IF neuron will always be bounded, i.e. $c\leq x \leq x_{thres}$, hence we do not need a semi-passivity argument to guarantee the solutions of such a model to be bounded.
\end{remark}

\section{Synchronization of diffusively coupled neuronal oscillators}\label{sec:semi-passivity-appl}
In the previous Section we showed that the solutions of diffusive coupled neurons (of the Hodgkin-Huxley, the Morris-Lecar, the FitzHugh-Nagumo and the Hindmarsh-Rose type) remain bounded. Using these results we provide conditions for which the neurons end up in synchrony.

Since the matrix $C B$ is nonsingular, the systems \eqref{eq:systems_sync} can be transformed into the following form
\beq\label{eq:systems_sync_normal_form}
    \splt{
        \dot{y}_j &= a(y_j,z_j)+ C B u_j= a(y_j,z_j)+ u_j, \\
        \dot{z}_j &= q(z_j,y_j),
    }
\eeq
where $y_j\in\Real$, $u_j\in\Real$, $z_j\in\Real^{m}$, $m=n-1$, and sufficiently smooth functions $a:\Real\times\Real^{m}\ra\Real$, $q:\Real^{m}\times\Real\ra\Real^{m}$.

\begin{thm}\cite{pog2001}\label{thm:sync}
Consider the $k$ systems \eqref{eq:systems_sync_normal_form} and assume that:
\ben
    \item each system
        \beq
            \splt{
                \dot{y}_j &= a(y_j,z_j)+ u_j, \\
                \dot{z}_j &= q(z_j,y_j),
            }
        \eeq
        is strictly semi-passive;
    \item there exists a $\Ccal^2$-smooth positive definite function $V_0:\Real^m \ra \Real_+$ and a positive number $\alpha \in \Real$ such that the following inequality is satisfied
        \beq\label{eq:thm1_eq2}
            \brac{\nabla V_0(z^\prime-z^{\prime\prime})}^\top\brac{q(z^\prime,y^\prime) - q(z^{\prime\prime},y^\prime)} \leq - \alpha \norm{z^\prime-z^{\prime\prime}}^2
        \eeq
        for all $z^{\prime},z^{\prime\prime}\in\Real^m$ and $y^{\prime} \in \Real$.
\een
Then, for all positive semi-definite matrices $\Gamma$ all solutions of the closed-loop system \eqref{eq:systems_sync_normal_form}, \eqref{eq:diffuse_coupling2} are ultimately bounded. Let the eigenvalues $\lambda_j$ of $\Gamma$ be ordered as $0=\lambda_1< \lambda_2 \leq \ldots\leq\lambda_k$. Then there exists a positive number $\bar{\lambda}$ such that if $\lambda_2 \geq \bar{\lambda}$ there exists a globally asymptotically stable subset of the diagonal set
\beq
    \Acal = \left\{ y_j \in \Real, z_j \in \Real^m : y_i = y_j, z_i = z_j, i,j = 1,\ldots,k \right\}.
\eeq
\end{thm}

\begin{remark}
One can easily verify that Theorem \ref{thm:sync} remains true in case that each system \eqref{eq:systems_sync_normal_form} is semi-passive in $\Dcal$, for $\Dcal$ as defined in Definition \ref{def:semi-passivity}.
\end{remark}

According to Theorem \ref{thm:sync} the problem of examining the asymptotic stability of the synchronized state of all oscillators in the network is reduced to
\ben
    \item verification of the assumptions for an individual oscillator, and
    \item computation of the eigenvalues of the coupling matrix $\Gamma$.
\een
It follows that if for a given network topology the coupling is large enough, i.e. $\lambda_2$ exceeds the threshold $\bar{\lambda}$, then the neurons will synchronize. Moreover, once the threshold value $\bar{\lambda}$ is known one can easily determine whether the neurons in networks with different topologies synchronize or not by computing the eigenvalues of the corresponding coupling matrix. This is the Wu-Chua conjecture \cite{wu1996}. The effect of the network topology on the synchronization can also be investigated using, for instance, the \emph{Connecting Graph Stability method} \cite{bel2005}.

\subsection{Convergent systems}
There exists a sufficient condition to check whether inequality \eqref{eq:thm1_eq2} of Theorem \ref{thm:sync} is satisfied or not. Therefore, let us introduce the notion of convergent systems.

\begin{defn}[Convergent systems]\cite{dem1967,pav2006}
Consider the system
\beq\label{eq:sys_conv}
\dot{z} = q(z,w(t)),
\eeq
where the external signal $w(t)$ is taking values from a compact set $\Wcal \subset \Real$.
The system \eqref{eq:sys_conv} is called \emph{convergent} if
\ben
    \item all solutions $z(t)$ are well-defined for all $t\in(-\infty,\ +\infty)$ and all initial conditions $z(0)$,
    \item there exists an unique globally asymptotically stable solution ${z}_w(t)$ on the interval $t\in(-\infty,\ +\infty)$ from which it follows
        \beq
        \lim \limits_{t\rightarrow\infty}\norm{ z(t) - {z}_w(t) } = 0
        \eeq
        for all initial conditions.
\een
\end{defn}
The long term motion of such systems is solely determined by the driving input $w(t)$ and not by initial conditions $z(0)$, i.e. the systems ``forget'' their initial conditions. A sufficient condition for a system to be convergent is presented in the next lemma.
\begin{lem}\cite{dem1967,pav2006}\label{lem:conv}
If there exists a positive definite symmetric $m \times m$ matrix $P$ such that all eigenvalues $\lambda_i(Q)$ of the symmetric matrix
\beq\label{eq:conv_test}
    Q(z,w) = \frac{1}{2}\left[P\left(\frac{\partial q}{\partial z}(z,w)\right) + \left(\frac{\partial q}{\partial z}(z,w)\right)^\top P\right]
\eeq
are negative and separated from zero, i.e. there is a $\delta > 0$ such that
\beq\label{eq:conv_test2}
    \lambda_i(Q)\leq -\delta < 0,
\eeq
with $i = 1,\ldots,m$ for all $z \in \Real^m$, $w \in \Wcal$, then the system \eqref{eq:sys_conv} is convergent.
\end{lem}
It follows that if there exists such a matrix $P$ such that each system $\dot{z}_j = q(z_j,y_j)$ satisfies \eqref{eq:conv_test}, \eqref{eq:conv_test2}, i.e. each system $\dot{z}_j = q(z_j,y_j)$ is convergent, then inequality \eqref{eq:thm1_eq2} of Theorem \ref{thm:sync} is satisfied.

One can easily verify that the internal dynamics of the models of Hodgkin-Huxley, Morris-Lecar, FitzHugh-Nagumo and \HR are convergent. (use $P = I$ in \eqref{eq:conv_test} and the result follows.)

\subsection{Illustrative examples} \label{sec:examples}
In the previous section we have shown that all four the models satisfy the semi-passivity condition. Moreover, the internal dynamics of these systems are equivalent to a convergent system. Therefore, according to Theorem \ref{thm:sync} a network consisting of the presented oscillators shows bounded solutions and, in case the coupling is strong enough, all oscillators will end up in perfect synchrony. However, the goal is here not to determine the exact threshold values for which the network starts to synchronize. Such threshold values can be expressed in terms of the system parameters (see, for instance \cite{oud2004} or \cite{bel2005} for \HR neurons), or they can be determined by computing, for instance, the transversal Lyapunov exponents of the coupled systems \cite{pec1998b}. Here, the goal is only to show that for large enough coupling the neurons will synchronize.

\subsubsection*{Synchronization of \HR oscillators}
\begin{figure}
\centering
\subfigure[Graph 1]
    {
        \begin{tikzpicture}[<->,auto,node distance=1.9cm,semithick]
            \tikzstyle{state}=[ball color=gray!30,circle,text=black]
                \node[state] (A) {$1$};
                \node[state] (B) [right of=A] {$2$};
                \node[state] (C) [below right of=B] {$3$};
                \node[state] (D) [below of=C] {$4$};
                \node[state] (E) [below left of=D] {$5$};
                \node[state] (F) [left of=E] {$6$};
                \node[state] (G) [above left of=F] {$7$};
                \node[state] (H) [above of=G] {$8$};
                \path (A) edge [bend left] node {} (B);
                \path (B) edge [bend left] node {} (C);
                \path (C) edge [bend left] node {} (D);
                \path (D) edge [bend left] node {} (E);
                \path (E) edge [bend left] node {} (F);
                \path (F) edge [bend left] node {} (G);
                \path (G) edge [bend left] node {} (H);
                \path (H) edge [bend left] node {} (A);
                \path (A) edge [bend right] node {} (C);
                \path (B) edge [bend right] node {} (D);
                \path (C) edge [bend right] node {} (E);
                \path (D) edge [bend right] node {} (F);
                \path (E) edge [bend right] node {} (G);
                \path (F) edge [bend right] node {} (H);
                \path (G) edge [bend right] node {} (A);
                \path (H) edge [bend right] node {} (B);
        \end{tikzpicture}
        \label{fig:sub:network_HR}
    }
\hskip 2ex
\subfigure[Graph 2]
    {
        \begin{tikzpicture}[<->,auto,node distance=1.9cm,semithick]
            \tikzstyle{state}=[ball color=gray!30,circle,text=black]
                \node[state] (A) {$1$};
                \node[state] (B) [right of=A] {$2$};
                \node[state] (C) [below right of=B] {$3$};
                \node[state] (D) [below of=C] {$4$};
                \node[state] (E) [below left of=D] {$5$};
                \node[state] (F) [left of=E] {$6$};
                \node[state] (G) [above left of=F] {$7$};
                \node[state] (H) [above of=G] {$8$};
                \path (A) edge [bend left] node {} (B);
                \path (A) edge node {} (E);
                \path (B) edge [bend left] node {} (C);
                \path (C) edge [bend left] node {} (D);
                \path (C) edge node {} (F);
                \path (C) edge node {} (G);
                \path (D) edge [bend left] node {} (E);
                \path (E) edge [bend left] node {} (F);
                \path (F) edge [bend left] node {} (G);
                \path (G) edge [bend left] node {} (H);
                \path (H) edge [bend left] node {} (A);
        \end{tikzpicture}
        \label{fig:sub:network_ML}
    }
    \caption{Eight diffusively coupled oscillators. Each interconnection has weight $\gamma$.}
\end{figure}
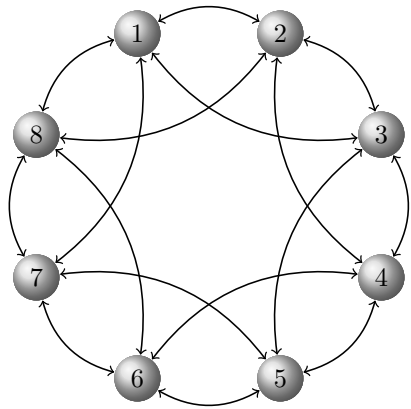
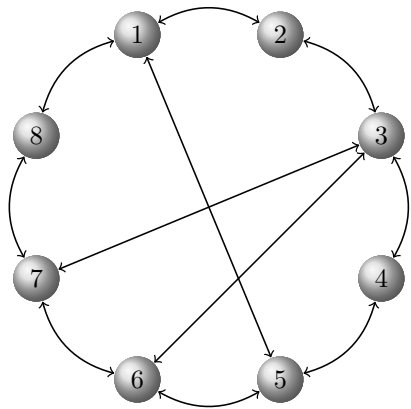

\begin{figure}
    \begin{center}
        \includegraphics[width=12cm]{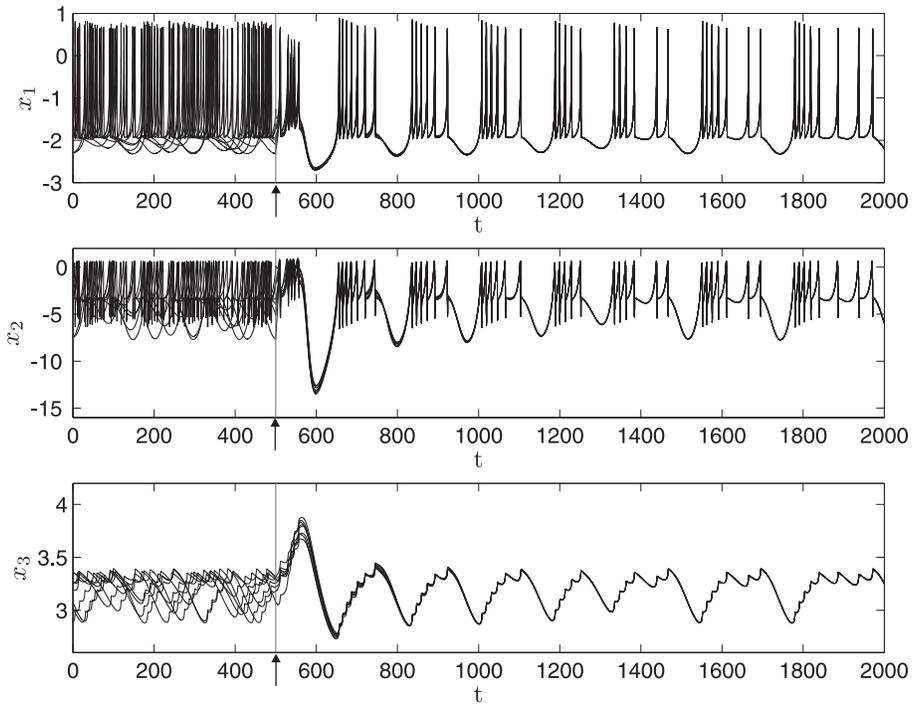}
        \caption{Synchronization of the eight \HR chaotic oscillators.}
        \label{fig:sync_HR}
    \end{center}
\end{figure}

Consider a network of eight diffusively coupled \HR neurons
\beq
    \splt{
    \dot{x}_{j,1} &= -a x_{j,1}^3 + b x_{j,1}^2 + x_{j,2} -x_{j,3} + I + u_j \\
    \dot{x}_{j,2} &= c-d x_{j,1}^2 - x_{j,2}\\
    \dot{x}_{j,3} &= r\brac{s \brac{x_{j,1} + w}-x_{j,3}}
    }
\eeq
where $j = 1,\ldots,8$ denotes the number of the oscillator in the network. We use the following set of parameters: $a=1,\ b=3,\ c=1,\ d=5,\ r=0.005,\ s=4,\ w=1.6180,\ I=3.25$. With these parameters each \HR neuron has chaotic solutions \cite{hin1984}. Let the eight oscillators be connected as shown in Figure \ref{fig:sub:network_HR} with corresponding coupling matrix

\beq
    \Gamma^1 = \bbm 4\gamma & -\gamma & -\gamma & 0 & 0 & 0 & -\gamma & -\gamma\\
                  -\gamma & 4\gamma & -\gamma & -\gamma & 0 & 0 & 0 & -\gamma\\
                  -\gamma & -\gamma & 4\gamma & -\gamma & -\gamma & 0 & 0 & 0\\
                  0 & -\gamma & -\gamma & 4\gamma & -\gamma & -\gamma & 0 & 0\\
                  0 & 0 & -\gamma & -\gamma & 4\gamma & -\gamma & -\gamma & 0\\
                  0 & 0 & 0 & -\gamma & -\gamma & 4\gamma & -\gamma & -\gamma\\
                  -\gamma & 0 & 0 & 0 & -\gamma & -\gamma & 4\gamma & -\gamma\\
                  -\gamma & -\gamma & 0 & 0 & 0 & -\gamma & -\gamma & 4\gamma \ebm
\eeq
The smallest nonzero eigenvalue of $\Gamma^1$ is $\lambda_2^1 \approx 2.58 \gamma$. Our simulations show that the neurons synchronize when $\gamma \geq 0.387$, which corresponds to $\bar{\lambda}^1 = 1.00$. (This agrees with the numerical results obtained in, for instance, \cite{bel2005}, where it is shown that two diffusively coupled \HR neurons synchronize when the coupling strength $\gamma\geq 0.50$, i.e. $\bar{\lambda} = 1.00$.)
Figure \ref{fig:sync_HR} shows the simulation results of the network of \HR oscillators with coupling $\gamma = 0.39$ such that $\lambda_2^1 \approx 1.01$. The top panel shows the $x_1$ states of the eight oscillators, the middle panel shows the $x_2$ states and the $x_3$ states are depicted in the bottom panel. The first $500\ [s]$ the systems are uncoupled and one sees the systems are not synchronized. After $500 \ [s]$ the coupling becomes active, indicated by the arrows in Figure \ref{fig:sync_HR}, and all systems rapidly synchronize.

\subsubsection*{Synchronization of \ML oscillators}
\begin{figure}
    \begin{center}
        \includegraphics[width=12cm]{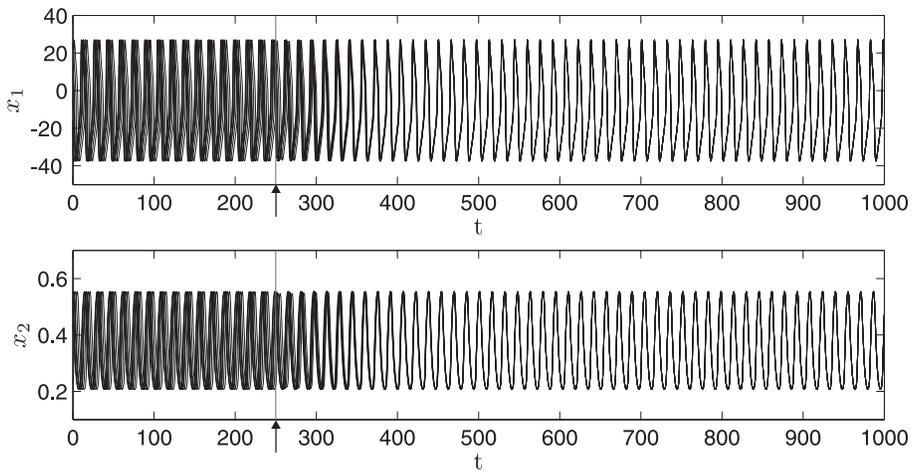}
        \caption{Synchronization of eight \ML oscillators.}
        \label{fig:sync_ML}
    \end{center}
\end{figure}

Next we synchronize eight \ML oscillators which are connected according to the graph depicted in Figure \ref{fig:sub:network_ML}. The corresponding coupling matrix is given as
\beq
    \Gamma^2 = \bbm 3\gamma & -\gamma & 0 & 0 & -\gamma & 0 & 0 & -\gamma\\
                  -\gamma & 2\gamma & -\gamma & 0 & 0 & 0 & 0 & 0\\
                  0 & -\gamma & 4\gamma & -\gamma & 0 & -\gamma & -\gamma & 0\\
                  0 & 0 & -\gamma & 2\gamma & -\gamma & 0 & 0 & 0\\
                  -\gamma & 0 & 0 & -\gamma & 3\gamma & -\gamma & 0 & 0\\
                  0 & 0 & -\gamma & 0 & -\gamma & 3\gamma & -\gamma & 0\\
                  0 & 0 & -\gamma & 0 & 0 & -\gamma & 3\gamma & -\gamma\\
                  -\gamma & 0 & 0 & 0 & 0 & 0 & -\gamma & 2\gamma \ebm
\eeq
such that the smallest nonzero eigenvalue of $\Gamma^2$ is $\lambda_2^2 \approx 1.27 \gamma$. Each \ML oscillator is given by the following set of equations
\beq
    \splt{
        C\dot{x}_{j,1} =& g_L\brac{E_L-x_{j,1}} + g_{Ca}\alpha_\infty\brac{x_{j,1}}\brac{E_{Ca}-x_{j,1}}+\\
        &+g_K x_{j,2} \brac{E_K-x_{j,1}}+ I + u_j,\\
        \dot{x}_{j,2} =& \eta\brac{x_{j,1}}\brac{\beta_\infty(x_{j,1}) - x_{j,2}},
    }
\eeq
with $j = 1,\ldots,k$ denoting the number of the oscillator in the network and functions
\beq
    \splt{
        \alpha_\infty(s) &= \frac{1}{2}\brac{1+\tanh\brac{\frac{s - E_1}{E_2}}}, \\
        \beta_\infty(s) &= \frac{1}{2}\brac{1+\tanh\brac{\frac{s - E_3}{E_4}}}, \\
        \eta(s) &= \bar{\eta}\cosh\brac{\frac{s-E_3}{2 E_4}}.
    }
\eeq
We used $C = 1$, $g_L=0.5$, $E_L=-50$, $g_{Ca}=1.1$, $E_{Ca}=100$, $g_K=2$, $E_K=-50$, $I= 30$, $E_1=-1$, $E_2=15$, $E_3=0$, $E_4=30$, $\bar{\eta}=5$ in our numerical simulations. Figure \ref{fig:sync_ML} shows the simulation results for the eight diffusively coupled \ML oscillators with $\gamma = 0.01$. The first $250 \ [s]$ the oscillators are uncoupled and do not synchronize. Then, after $250 [s]$ the coupling is turned on, which is again indicated by the arrow, and all oscillators become synchronized.

\section{Diffusion driven instabilities}\label{sec:instabilities}

In this section we show using two simple examples that it is not trivial that systems interacting via diffusive coupling have bounded solutions and possibly end up in synchrony. In particular, we demonstrate that diffusive coupling 1) can make the solutions of the interconnected systems to become unbounded, and 2) can make systems, which have an asymptotically stable equilibrium in isolation, to produce stable oscillations.

\begin{exam}[Unbounded solutions]\label{exam:diff_unb}
Consider the linear (non-minimum phase\footnote{a system is non-minimum phase if it has unstable zero dynamics, i.e. the internal dynamics with constraint $y(t) = 0$ are unstable, cf \cite{pog1999}.}) stable transfer function
\beq
    H(s) = \frac{s^2-s+1}{s^3+2s^2+2s+1}.
\eeq
A possible state space realization for the system is
\beq \label{eq:sys_unstable}
    \dot{x} = Ax+Bu,\quad y =C x,
\eeq
where
\beq
    A = \bbm 1& -1& 1 \\ 1 & 0& 0 \\ -4 &2 & -3 \ebm,\quad B = C^\top = \bbm 0 \\0 \\ 1 \ebm.
\eeq
Consider now two diffusively coupled systems \eqref{eq:sys_unstable}
\beq
    \splt{
        \dot{x}_1 = A x_1 + \gamma BC(x_2-x_1), \\
        \dot{x}_2 = A x_2 + \gamma BC(x_1-x_2), \\
    }
\eeq
Clearly the origin of each uncoupled system is globally asymptotically stable. However, the system is not semi-passive, and when $\gamma>0.6512$ (for $\gamma = 0.6512$ the system undergoes a Poincar\'e-Andronov-Hopf bifurcation \cite{pog1999}) the solutions of the interconnected systems become unbounded.
\end{exam}

Example \ref{exam:diff_unb} shows how the diffusive coupling between two not semi-passive systems results in unbounded solutions. A similar phenomena is encountered in networks of diffusively coupled Chua circuits, cf. \cite{steen2006}. The piecewise linear model of the Chua circuit is not semi-passive (the Chua attractor is not globally stable) and due to the interaction the trajectories of the systems can be driven outside the domain of attraction such that the solutions grow unbounded.

\begin{figure}
    \begin{center}
        \includegraphics[width=12cm]{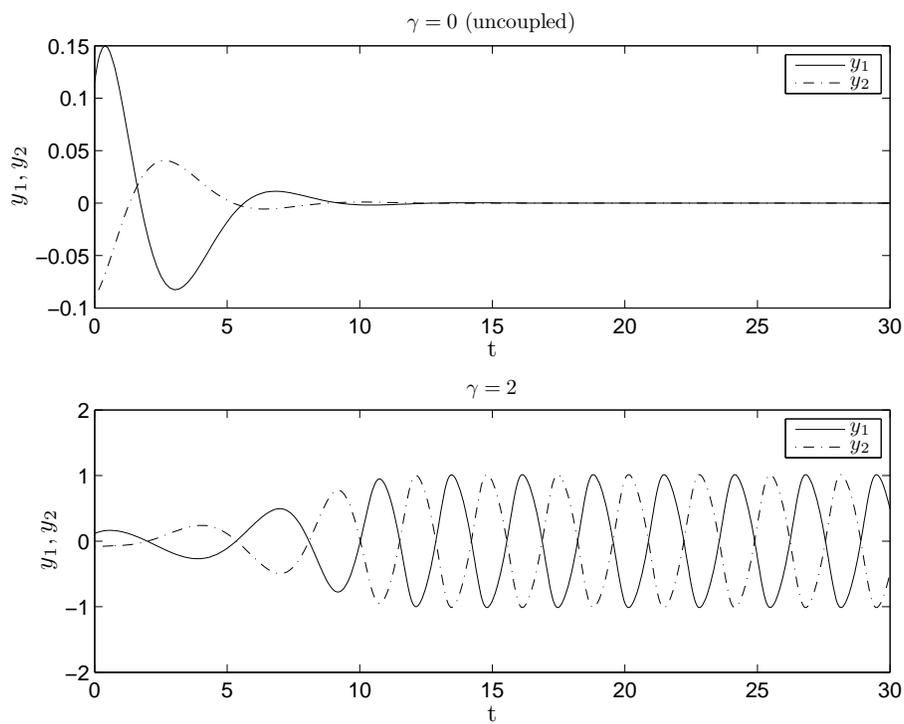}
        \caption{Diffusion driven oscillations.}
        \label{fig:example_osc}
    \end{center}
\end{figure}

The following example is taken from \cite{pog1999}. It shows how two systems, which both have an asymptotically stable equilibrium in absence of interaction, start to produce stable oscillations when the systems interact via diffusive coupling.
\begin{exam}[Diffusion driven oscillations]
Consider two systems which interact via diffusive coupling:
\beq\label{eq:sys_unstable_osc}
    \splt{
        \dot{x}_1 = A x_1(1+\norm{x_1}^2) + \gamma BC(x_2-x_1), \\
        \dot{x}_2 = A x_2(1+\norm{x_2}^2) + \gamma BC(x_1-x_2), \\
    }
\eeq
where matrices $A$, $B$ and $C$ are as presented above. Again the origin of an isolated system is asymptotically stable and when $\gamma = 0.6512$ the (linearized) system undergoes a Poincar\'e-Andronov-Hopf bifurcation. Hence the coupled systems \eqref{eq:sys_unstable_osc} start to produce stable oscillations whenever $\gamma>0.6512$, see Figure \ref{fig:example_osc} for simulation results.
\end{exam}
The key mechanism for the oscillations is the Poincar\'e-Andronov-Hopf bifurcation and the non-minimum phaseness of the systems.
The diffusive interaction between initially silent cells is essential for generating stable oscillatory behavior in some neuronal (and other biological) systems, see \cite{loe2001} and the references therein. The authors demonstrate that the main reason for the oscillations is that the internal variables, e.g. (in)activation particles, have the tendency to oscillate. However, these oscillations are being suppressed through a negative feedback mechanism. The diffusive coupling will destroy the feedback mechanism causes the internal variables and, hence, the membrane potential to start to oscillate. Note that the mechanism is the same as in our example, i.e. the internal dynamics are not minimum phase. However, the goal here is not to discuss the machinery for the generation of these oscillations in detail. We refer the reader to \cite{pog1999} for more details.

Note that the four models described above do have minimum phase internal dynamics since the internal dynamics are convergent. Hence no ``spontaneous'' oscillations due to diffusive interaction will occur in networks of Hodgkin-Huxley, Morris-Lecar, FitzHugh-Nagumo and \HR neurons.

\section{Discussion}\label{sec:discussion}
We have presented sufficient conditions for synchronization in networks of diffusively coupled neuronal oscillators. The results are constructive in the following sense:
\ben
    \item we have considered different classes of neuronal oscillators, i.e. neuronal oscillators whose dynamics are described via the \HH formalism and neuronal oscillators of the \FN and \HR type;
    \item we have shown that all these oscillators are semi-passive with a quadratic storage function. A consequence is that, when semi-passive systems are being coupled via coupling of the type \eqref{eq:diffuse_coupling}, the network possesses bounded solutions (see Proposition \ref{prop:bounded}).
    \item the internal dynamics of all these neuronal oscillators are convergent, i.e. the subsystem $\dot{z} = q(z,y)$ of each oscillator satisfies the conditions as stated in Lemma \ref{lem:conv}. As a result \eqref{eq:thm1_eq2} of Theorem \ref{thm:sync} is satisfied;
    \item since the oscillators are semi-passive and the internal dynamics are convergent it is possible, according to Theorem \ref{thm:sync}, that all oscillators in the network end up in stable synchrony. The criteria for synchronization is that for a given network topology the strength of the interconnections is large enough since topology and coupling strength influence the smallest nonzero eigenvalue of the coupling matrix $\Gamma$.
\een
Theorem \ref{thm:sync} allows to decompose the problem of finding (sufficient) conditions for stable synchronization in the network of $k$ coupled oscillators into some conditions of the individual oscillators (semi-passivity and internal convergent dynamics) and conditions on the network (topology and coupling strength, which both influence the smallest nonzero eigenvalue of the coupling matrix). Our results can therefore be applied to neuronal networks interconnected via strong coupling and with general network topology. Our theory supports the result of \cite{lab2003} and the simulation results of Chow and Kopell \cite{cho2000} for strong coupling (note that the model of the interneuron and the Traub-Miles model are both semi-passive and have convergent internal dynamics). Moreover, our results hold even when the neurons behave chaotically.
In \cite{erm2009} the authors discuss the emergence of clusters in all-to-all coupled networks as function of the coupling strength. Those clusters might emerge when the coupling is not strong enough to end up in synchrony. The emergence of clusters for diffusively coupled neurons satisfying the assumptions of Theorem \ref{thm:sync} can be explained for general network topology using the theory discussed in \cite{pog2002,pog2008}.
The goal of this paper is to show that neurons interconnected via gap junctions will posses bounded solutions and, moreover, synchronize whenever the coupling strength is large enough. Determining sharp synchronization thresholds however will still depend highly on the type of neurons involved and their specific set of parameters.
We have also shown that diffusively coupled systems which are not semi-passive might have unbounded solutions. A probably more interesting property, at least from the biological point of view, is that diffusively coupled non-minimum phase systems which are initially silent can start to produce stable oscillations.

In this paper we considered networks of diffusively coupled neurons without any time-delay. However, in a physical system one would expect that it takes some (small amount of) time to transmit a signal. Therefore it is interesting to analyse synchronization in networks where time-delays are included in the coupling. Sufficient conditions for synchronization in time-delayed networks are presented in, for instance, \cite{ogu2008}. Here, the semi-passivity property in combination with a small-gain theorem provides a sufficient condition for boundedness of the trajectories of the coupled systems. Next sufficient conditions for synchronization in terms of Linear Matrix Inequalities (LMIs) are derived. However, solving the LMIs is computationally involving, especially when the networks become large and complicated. Moreover, the results might be very conservative. 
It would be interesting as well to investigate the emergence of stable synchronization in pulse-coupled networks. This is because most neurons are actually coupled via so-called chemical synapses, i.e. an impulsive type of coupling. It is shown in \cite{mir1990} that certain \IF neurons in a network with all-to-all connections synchronize for almost any initial condition. In \cite{som1993,som1995,rub2002} synchronization of more realistic neuronal oscillators in pulse-coupled networks is discussed. However, rigorous constructive results about what conditions the oscillators should satisfy and the effect of a particular network topology on the synchronization are not present nowadays. Even for systems that can be represented as the seemingly simple (pulse-)coupled Kuramoto oscillators, cf \cite{kur1991}, the problem of global synchronization is not tackled in full generality. Networks with strong interactions and/or chaotic regimes remain problematic. It is for future research to explore the possibilities in these topics.

\bibliographystyle{elsarticle-harv}
\bibliography{mybib}

%

\end{document}